\documentstyle[11pt,amssymb]{article}
\begin{document}

\newcommand{\nnl}{\nl[6mm]}
\newcommand{\enl}{\\[6mm]}
\newcommand{\nllbnl}[1]{\nl[-3mm]\label{#1}\\[-3mm]}
\newcommand{\nle}{\nl[-3mm]\\[-3mm]}
\newcommand{\nl}{\nonumber\\}
\newcommand{\bl}{&&\quad}
\newcommand{\ab}{\allowbreak}

\newcommand{\iint}{{\int\hskip-3mm\int}}
\newcommand{\iiint}{{\int\hskip-3mm\int\hskip-3mm\int}}
\newcommand{\dddot}[1]{{\mathop{#1}\limits^{\vbox to 0pt{\kern 0pt
\hbox{.{\kern-0.25mm}.{\kern-0.25mm}.}\vss}}}}
\renewcommand{\leq}{\leqslant}
\renewcommand{\geq}{\geqslant}

\renewcommand{\theequation}{\thesection.\arabic{equation}}
\let\ssection=\section
\renewcommand{\section}{\setcounter{equation}{0}\ssection}

\newcommand{\be}{\begin{equation}}
\newcommand{\ee}{\end{equation}}
\newcommand{\bes}{\begin{eqnarray}}
\newcommand{\ees}{\end{eqnarray}}
\newcommand{\eens}{\nonumber\end{eqnarray}}

\renewcommand{\/}{\over}
\renewcommand{\d}{\partial}
\newcommand{\ddt}[1]{{\d{#1}\/\d t}}
\newcommand{\dNx}{d^N\!x\ }
\newcommand{\dNy}{d^N\!y\ }
\newcommand{\summ}[1]{{\sum_{|\mm|\leq #1}}}

\newcommand{\eps}{\epsilon}
\newcommand{\gm}{\gamma}
\newcommand{\dlt}{\delta}
\newcommand{\th}{\theta}
\newcommand{\al}{\alpha}
\newcommand{\bt}{\beta}
\newcommand{\si}{\sigma}
\newcommand{\ka}{\kappa}
\newcommand{\la}{\lambda}
\newcommand{\rep}{\varrho}

\newcommand{\bb}{{\frak b}}
\newcommand{\cc}{{\frak c}}
\newcommand{\oj}{{\frak g}}
\newcommand{\hh}{{\frak h}}

\newcommand{\xmu}{\xi^\mu}
\newcommand{\xnu}{\xi^\nu}
\newcommand{\xrho}{\xi^\rho}
\newcommand{\xsi}{\xi^\si}
\newcommand{\ynu}{\eta^\nu}
\newcommand{\dmu}{\d_\mu}
\newcommand{\dnu}{\d_\nu}
\newcommand{\dsi}{\d_\si}
\newcommand{\dtau}{\d_\tau}
\newcommand{\drho}{\d_\rho}
\newcommand{\qmu}{q^\mu}
\newcommand{\qnu}{q^\nu}
\newcommand{\qrho}{q^\rho} 
\newcommand{\qsi}{q^\si}
\newcommand{\qtau}{q^\tau}
\newcommand{\pmu}{p_\mu}
\newcommand{\pnu}{p_\nu}

\newcommand{\fa}{\phi_\alpha}
\newcommand{\fb}{\phi_\beta}
\newcommand{\pa}{\pi^\alpha}
\newcommand{\pb}{\pi^\beta}
\newcommand{\fsa}{\phi^{*\alpha}}
\newcommand{\fsb}{\phi^{*\beta}}
\newcommand{\psa}{\pi^*_\alpha}
\newcommand{\psb}{\pi^*_\beta}
\newcommand{\Ea}{\EE^\alpha}
\newcommand{\Eb}{\EE^\beta}
\newcommand{\Eam}{\Ea_{,\mm}}

\newcommand{\fm}[1]{\phi_{#1,\mm}}
\newcommand{\fn}[1]{\phi_{#1,\nn}}
\newcommand{\pim}[1]{\pi^{#1,\mm}}
\newcommand{\pin}[1]{\pi^{#1,\nn}}
\newcommand{\fsm}[1]{\phi^{*#1}_{,\mm}}
\newcommand{\psm}[1]{\pi^{*,\mm}_{#1}}

\newcommand{\hphi}{{\hat\phi}}

\newcommand{\tpi}{{1\/2\pi i}}
\newcommand{\half}{{1\/2}}
\newcommand{\Np}[1]{{N+p\choose p #1}}

\newcommand{\mm}{{\mathbf m}}
\newcommand{\nn}{{\mathbf n}}
\newcommand{\mmu}{{\underline \mu}}

\newcommand{\LL}{{\cal L}}
\newcommand{\DD}{{\cal D}}
\newcommand{\EE}{{\cal E}}
\newcommand{\FF}{{\cal F}}
\newcommand{\GG}{{\cal G}}
\newcommand{\PP}{{\cal P}}
\newcommand{\QQ}{{\cal Q}}
\newcommand{\KK}{{\cal KT}}
\newcommand{\NS}{{\cal N}_S}
\newcommand{\BB}{{\cal B}}

\newcommand{\CF}{C^\infty}
\newcommand{\QKT}{Q_{KT}}

\newcommand{\Lxi}{\LL_\xi}
\newcommand{\Leta}{\LL_\eta}

\newcommand{\into}{\hookrightarrow}
\newcommand{\e}{{\rm e}}
\newcommand{\tdiff}{{\widetilde{diff}}}

\newcommand{\bra}[1]{\big{\langle}#1\big{|}}
\newcommand{\ket}[1]{\big{|}#1\big{\rangle}}
\newcommand{\no}[1]{{\,:\kern-0.7mm #1\kern-1.2mm:\,}}

\newcommand{\RR}{{\Bbb R}}
\newcommand{\CC}{{\Bbb C}}
\newcommand{\ZZ}{{\Bbb Z}}

\topmargin 1.0cm

\vspace*{-3cm}
\pagenumbering{arabic}
\begin{flushright}
{\tt gr-qc/9909039}
\end{flushright}
\vspace{12mm}
\begin{center}
{\huge The diffeomorphism algebra approach to quantum gravity}\\[14mm]
\renewcommand{\baselinestretch}{1.2}
\renewcommand{\footnotesep}{10pt}
{\large T. A. Larsson\\
}
\vspace{12mm}
{\sl Vanadisv\"agen 29\\
S-113 23 Stockholm, Sweden}\\
email: tal@hdd.se
\end{center}
\vspace{3mm}
\begin{abstract}
The representation theory of non-centrally extended Lie algebras of
Noether symmetries, including spacetime diffeomorphisms and 
reparametrizations of the observer's trajectory, has recently
been developped. It naturally solves some long-standing problems
in quantum gravity, e.g. the role of diffeomorphisms and the causal 
structure, but some new questions also arise.
\end{abstract}
\renewcommand{\baselinestretch}{1.5}

\section{Introduction}
In overviews of conceptual problems in quantum gravity
\cite{Ish91,Rov98}, the role of spacetime diffeomorphisms is often
listed as one of the major difficulties. To me, it appears obvious
that this issue can not be settled before the representation theory
of $diff(N)$, the diffeomorphism algebra in $N$ spacetime dimensions,
has been worked out. The classical (proper) representations are given by 
tensor densities; this was formally proven by Rudakov in 1974
\cite{Rud74} (see also \cite{ER96}), 
but it has of course been implicitly known for a century.
However, in quantum theory one is interested in projective 
representations, i.e. group representations up to a (local) phase. 
On the Lie algebra level, this corresponds to an abelian but non-central
(except when $N=1$) extension of $diff(N)$.
In one dimension, the problem has been long solved. The abelian (even
central) extension of $diff(1)$ is the Virasoro algebra, tensor densities
are usually called primary fields, and the interesting quantum modules
are of lowest-energy type \cite{BPZ84,FF82,FMS96,GO86,Kac79}. 

During the nineties, the higher-dimensional analogs of the extensions 
\cite{Dzhu96,ERM94,Lar89,Lar91}
and the representation theory \cite{BB98,Bil97,ERM94,Lar97,Lar98,Lar99}
have largely been worked out. Related work for gauge (or current)
algebra was previously carried out in \cite{EMY92,MEY90}.
In particular, in \cite{Lar98} I discovered the geometric picture
underlying these representations. The starting point is the algebra
$diff(N)\oplus diff(1)$, where the extra $diff(1)$ factor describes
reparametrizations of the observer's trajectory. One expands all
fields in a multi-dimensional Taylor series around the observer's
present position and discard Taylor coefficients of order higher than
$p$ ($p$ finite), i.e. one passes to $p$-jet space. Then, one
introduces canonically conjugate objects (jet momenta), normal orders
and representents the resulting Heisenberg algebra on a unique Fock space.
The point is that this procedure, which is reviewed in detail in 
section 2, results in a well-defined abelian extension of 
$diff(N)\oplus diff(1)$, called the DRO (Diffeomorphism, 
Reparametrization, Observer) algebra $DRO(N)$.

This result should be contrasted with the obvious 
generalization of the one-dimensional construction:
start from tensor densities, introduce canonical momenta, and
normal order with respect to an arbitrarily chosen time direction.
Such a na\"\i ve approach can not work, since infinities (proportional 
to the number of time-independent functions) arise.

In the recent work \cite{Lar99}, which is reviewed in section 3, I
introduced dynamics into the picture. There are two reasons for doing 
so: mathematically, one wishes to reduce the representations in 
section 2; physically, any theory of gravity must contain information
about the Einstein and geodesic equations. Technically, I introduce
a generalized Koszul-Tate complex \cite{HT92,Sta97}, involving not
only fields and anti-fields but also their canonical momenta. After
projection to jet space, this construction survives quantization,
and the cohomology groups are $DRO(N)$ modules.

In section 4 I discuss in what sense that DRO algebra representations
constitute a theory of quantum gravity. Of course, it tautologically 
solves the problem with diffeomorphisms, but some other questions 
also acquire natural answers. The parameter along the observer's 
trajectory defines a natural causal structure and also provides a
unique definition of the vacuum, as being annihilated by all negative
frequency modes. Unfortunately, the present formalism also introduces 
some new problems. There are some technicalities, such as
the limit $p\to\infty$, but the main open question is how to recover 
standard quantum results in nice cases, e.g. flat metric. 

In the section 5, I discuss some of the competition on the quantum
gravity marketplace, notably string
theory, from the point of view of (super-)diffeomorphisms. 
Finally, I argue in section 6 that the DRO algebra can be useful in
another application: the classification of three-dimensional critical
exponents.

\section{DRO algebra}
This section is based on \cite{Lar98}.
$DRO(N)$ is a four-parameter abelian but non-central extension of
$diff(N)\oplus diff(1)$ by the commutative algebra of local functionals
in the observer's spacetime trajectory. Explicitly, it is the
Lie algebra defined by the brackets
\bes
[\Lxi,\Leta] &=& \LL_{[\xi,\eta]} 
+ \tpi\int dt\ \dot\qrho(t) 
\Big( c_1 \drho\dnu\xmu(q(t))\dmu\ynu(q(t)) +\nl
\bl+ c_2 \drho\dmu\xmu(q(t))\dnu\ynu(q(t)) \Big), \nl
{[}L_f, \Lxi] &=& {c_3\/4\pi i} \int dt\ 
(\ddot f(t) - i\dot f(t))\dmu\xmu(q(t)), \nl
{[}L_f,L_g] &=& L_{[f,g]} 
+ {c_4\/24\pi i}\int dt (\ddot f(t) \dot g(t) - \dot f(t) g(t)), 
\label{DRO}\\
{[}\Lxi,\qmu(t)] &=& \xmu(q(t)), \nl
{[}L_f,\qmu(t)] &=& -f(t)\dot\qmu(t), \nl
{[}\qmu(s), \qnu(t)] &=& 0,
\eens
continued by linearity and Leibniz' rule. Here $\xi=\xmu(x)\dmu$, 
$x\in\RR^N$, and $\eta$ are spacetime vector fields, and
$\qmu(t)$, $t\in S^1$, $\mu = 0,1, \ldots N-1$ describes the 
observer's trajectory in spacetime. 

In our convention, $gl(N)$ has basis $T^\mu_\nu$ and brackets
\be
[T^\mu_\nu,T^\si_\tau] = 
\dlt^\si_\nu T^\mu_\tau - \dlt^\mu_\tau T^\si_\mu.
\label{glN}
\ee
Denote the matrix elements of a $gl(N)$ representation $\rep$ by 
$\rep^\al_\bt(T^\mu_\nu)$.
The classical representations of $DRO(N)$ (with all ``abelian charges'' 
$c_1 = c_2 = c_3 = c_4 = 0$) are given by a set of fields over 
spacetime $\fa(x)$, where $x\in\RR^N$ and $\alpha$ labels different
tensor and internal components, and the observer's trajectory in 
spacetime $\qmu(t) \in \RR^N$, $t\in S^1$.
\bes
{[}\Lxi,\fa(x)] &=& -\xmu(x)\dmu\fa(x) 
- \dnu\xmu(x)\rep^\bt_\al(T^\nu_\mu) \fb(x), \nl
{[}L_f,\fa(x)] &=& 0, \nle
{[}\Lxi,\qmu(t)] &=& \xmu(q(t)), \nl
{[}L_f,\qmu(t)] &=& -f(t)\dot\qmu(t),
\nnl
{[}\Lxi,\pa(x)] &=&  -\xmu(x)\dmu \pa(x) 
+ \dnu\xmu(x)\pb(x) \rep^\al_\bt(T^\nu_\mu), \nl
{[}L_f,\pa(x)] &=& 0, \nle
{[}\Lxi,\pnu(t)] &=& -\dnu\xmu(q(t)) \pmu(t), \nl
{[}L_f,\pnu(t)] &=& -f(t)\dot\pnu(t) - \dot f(t) \pnu(t), 
\eens
We now expand all fields in a multi-dimensional Taylor series around 
the observer's present position.
\be
\fa(x) = \sum_{|\mm|\geq0} {1\/\mm!} \fm\al(t)(x-q(t))^\mm,
\label{jet}
\ee
where $\mm$ is a multi-index with length 
$|\mm| = \sum_{\mu=0}^{N-1} m_\mu$. 
The jets (Taylor coefficients) $\fm\al(t)$ depend on $t$ although
the field $\fa(x)$ does not, since this dependence enters through 
the expansion point. 
We can now construct the corresponding Fock space $J^p\FF$ by 
introducing
the jet momenta $\pmu(t)$ and $\pim\al(t)$, expand everything
(trajectory, jets, and all momenta) in a Fourier series in $t$, 
and proclaim that the Fock vacuum be
annihilated by the negative frequency modes. 
The main result of \cite{Lar98} is the explicit description of the
DRO algebra action on $J^p\FF$. Theorems 5.1 and
6.2 in that paper prove that the following operators 
provide a realization of the $DRO(N)$:
\bes
\Lxi &=& \int dt\ \no{\xmu(q(t)) \pmu(t)} - \sum_{|\mm|\leq|\nn|\leq p}
  \rep^\al_\bt(T^\mm_\nn(\xi(q(t)))) \no{ \fm\al(t)\pin\bt(t)  }, \nle
L_f &=& \int dt\  f(t)( - \no{\dot\qmu(t)\pmu(t)}
  -  \summ{p} \no{ \dot\fm\al(t)\pim\al(t)} )\nl
&&+ \la \summ{p} \no{ \fm\al(t)\pim\al(t)} ),
\eens
where $\la$ is a parameter (the ``causal weight'') and 
$T^\mm_\nn(\xi)$ are certain operators.
Normal ordering (indicated by double dots) is carried out by
moving negative frequency modes to the right. 
The values of the abelian charges $c_1$ -- $c_4$,
which depend on the data $p$, $\rep$, and $\la$, were also computed
in \cite{Lar98}

\section{Dynamics}
This section is based on \cite{Lar99}.
Let $S=\int \dNx \sqrt{|g(x)|} \LL(x)$ be an action with Lagrangian
$\LL(x)$. The solution to the Euler-Lagrange (EL) equations,
\be
\Ea(x) \equiv [\pa(x), S] = 0,
\label{EL}
\ee
defines the {\em stationary surface} $\Sigma$. 
Introduce an antifield $\fsa(x)$ with momentum $\psa(x)$, of
opposite Grassman parity compared to $\fa(x)$.
The fermionic Koszul-Tate (KT) generator,
\be
\QKT^{(1)} = \int \dNx \Ea(x) \psa(x),
\label{QKT1}
\ee
is nilpotent because $\Ea(x)$ commutes with the antifield momentum. 
For each Noether identity, of the form
\be
r^a(x) \equiv \int \dNy r^a_\al(x,y)\Ea(y)
= \int \dNy (-)^\al \Ea(y)r^a_\al(x,y) \equiv 0,
\label{Noether}
\ee
introduce a bosonic Noether antifield $\bb^a(x)$ with momentum 
$\cc_a(x)$. A new term has to be added to the KT generator (\ref{QKT1})
\be
\QKT^{(2)} = \iint \dNx \dNy (-)^\al r^a_\al(x,y) \fsa(y) \cc_a(x).
\label{Q2}
\ee
The cohomology of $\QKT = \QKT^{(1)} + \QKT^{(2)}$ is isomorphic to the
space of differential forms on the stationary surface $\Sigma$.

The Taylor coefficents depend on the parameter $t$ although the field
itself does not, because the expansion point $\qmu(t)$ does.
On the other hand, the RHS of (\ref{jet}) actually defines a function
$\fa(x,t)$ of two variables. To resolve this paradox we must impose
the condition ${\d \fa(x,t)/\d t} = 0$, which is equivalent to
\be
\DD_{\al,\mm}(t) \equiv
\dot\fm\al(t) - \dot\qmu(t)\phi_{\al,m+\mmu}(t) \approx 0.
\label{DD1}
\ee
The contribution to the KT generator is
\be
\QKT^{(\DD)} = \summ{p-1} \int dt\ \DD_{\al,\mm}(t) \gm^{\al,\mm}(t).
\ee
where an antifield $\bt_{\al,\mm}(t)$, with momentum 
$\gm^{\al,\mm}(t)$, was introduced.

Since the EL constraint (\ref{EL}) is a local functional, it can be
expanded in a Taylor series,
\be
\Ea(x) = \sum_{|\mm|\geq0} {1\/\mm!} \Eam(t)(x-q(t))^\mm.
\ee
The EL constraint now takes the form 
\be
\Eam(t) \approx 0, \qquad
\forall\, |\mm| \leq p - o_\al,
\label{ELm}
\ee
where $o_\al$ is the order of the EL equation.
It is implemented in cohomology by the KT generator
\be
\QKT^{(\EE)} = \summ{p-o_\al} \int dt\ \Eam(t) \psm\al(t),
\ee

However, the constraints (\ref{DD1}) and (\ref{ELm}) are not
independent, because
$\dot\Eam(t) - \dot\qmu(t)\Ea_{,\mm+\mmu}(t) \approx 0$.
Hence
\be
\BB^\al_{,\mm}(t) = \dot\fsm\al(t) - \dot\qmu(t)\fsa_{,\mm+\mmu}(t)
\ee
generates unwanted cohomology, which must be killed by new antifields; 
call these $\bt^{*\al}_{,\mm}(t)$ and their momenta
$\gm^{*,\mm}_\al(t)$. The contribution to the KT generator is
\be
\QKT^{(\BB)} = \summ{p-o_\al-1} \int dt\ 
\BB^\al_{,\mm}(t) \gm^{*,\mm}_\al(t).
\ee

Define the geodesic operator
\be
\GG_\mu(t) = g_{\mu\nu}(t) ( \ddot\qnu(t) + \Gamma(t) \dot\qnu(t) 
+ \Gamma^\nu_{\si\tau}(t)\dot\qsi(t)\dot\qtau(t) ) \approx 0,
\label{GG}
\ee
where $g_{\mu\nu}(t)$ and $\Gamma^\nu_{\si\tau}(t)$ are the zero-jets
corresponding to the metric and Levi-Civit\`a connection, and 
$\Gamma(t)$ is the connection for reparametrizations. To (\ref{GG})
corresponds the antifield (or ``anti-trajectory'') $q^*_\mu(t)$
with momentum $p^{*\mu}(t)$, and the KT charge
\be
\QKT^{(\GG)} = \int dt\ \GG_\mu(t) p^{*\mu}(t).
\ee

The total KT charge is $\QKT = \QKT^{(\DD)} + \QKT^{(\EE)} + \QKT^{(\BB)}
+\QKT^{(\GG)}$, plus further contributions from the Noether identities.

Define a physical state $\ket{phys} \in J^p\FF$ as a state that is
annihilated by the total KT generator, $\QKT\ket{phys} = 0$.
The state cohomology $H^\bullet_{state}(\QKT, J^p\FF)$
is the space of physical states modulo relations 
$\ket{phys} \sim \ket{phys} + \QKT\ket{}$. Cleary, the cohomology
groups $H^g_{state}(\QKT, J^p\FF)$ have two important properties:
\begin{enumerate}
\item
They are well defined $DRO(N)$ modules, for every choice of 
EL equations (\ref{EL}). This is so because the underlying space
$J^p\FF$ is well defined, the KT generator $\QKT$ commutes with 
$DRO(N)$, and $\QKT$ is not affected by normal ordering.
\item
They are obtained from the classical theory (dual to the EL equations)
by a quantization-like procedure.
\end{enumerate}

\section{Quantum gravity}
The representation theory described in the two previous is 
mathematically interesting in its own right, but how does it relate to
physics? Classically, the KT cohomology describes differential forms
on the stationary surface, so it contains information about classical
dynamics, i.e. the EL equations. ``Quantization'' then proceeds in four
steps: 
\begin{enumerate}
\item
Expand all fields in a Taylor series around the observer's present
position and truncate after terms of order $p$.
\item
Replace Poisson brackets by graded commutators everywhere.
\item
Normal order to remove infinities.
\item
Represent the resulting Heisenberg algebra on the unique Fock space
defined by postulating that negative frequency modes annihilate the
vacuum.
\end{enumerate}
It is not clear that this recipe is equivalent to the usual meaning
of quantization, although it is certainly closely related to it.
An indicatition that we really are dealing with a genuine quantum
theory is that the classical symmetry algebra $diff(N)\oplus diff(1)$
is represented projectively, i.e. it acquires a non-trivial abelian
extension.

{F}rom one point of view, the gravitational field is not very special;
it is just one of the fields $\fa(x)$ in the Lagrangian, which is 
quantized in the prescribed fashion.
However, it is somewhat special in the sense that it
also appears in the geodesic equation, which is needed to eliminate
the observer's trajectory.

Some common conceptual difficulties with other approaches to quantum
gravity are naturally resolved.
\begin{enumerate}
\item
All operators (jets, the trajectory, and their momenta) depend on 
``parameter time'', $t$. Its numerical value has no meaning, since
the Fock modules carry representations of the
reparametrization algebra, but it defines a natural causal structure.
\item
In order to define the vacuum and equal-time commutators in
standard canonical quantization, spacetime must be split into space 
and time. Such a decomposition clearly introduces a space-time
assymmetry, which greatly endangers general covariance of the
final results. In contrast, I define vacuum and equal-time commutators
with respect to the additional parameter $t$, while maintaining
manifest general covariance. Space-time symmetry could be broken also
within the present formalism, by introducing the constraints 
$L_f\approx0$, $q^0(t)\approx t$ \cite{Lar98}, but it is not mandatory.
\item
A common source of confusion is the claim that diffeomorphism symmetry
implies that all correlation functions vanish. This is necessarily true
only in the absense of abelian charges, but need not be true for the
$DRO$ algebra. In particular, the abelian extension of $diff(1)$ is 
the Virasoro algebra, which certainly can coexist with non-trivial
correlation functions.
\end{enumerate}

All other formulations of physics, both classical and quantum, can
classified as either Lagrangian or Hamiltonian. The present formulation,
however, is both. It is Hamiltonian in the sense that canonical momenta
and Poisson brackets/commutators are introduced, but Lagrangian in the
sense that all spacetime directions are treated on an equal footing.
A related point is that I keep velocities and momenta as independent
objects, instead of eliminating one of them by a relation of the form
\be
\pa = {\d(\sqrt{|g|} \LL)\/\d\d_0\phi}.
\ee

There are two difficulties of a technical nature \cite{Lar99}:
\begin{enumerate}
\item
There is an ambiguity in defining the constraints. 
If $\chi \approx 0$ defines the stationary surface, $A\chi \approx 0$
defines the same surface for every invertible operator $A$. However, 
although the surfaces defined by the two constraints are the same,
the operators $\chi$ and $A\chi$, and hence the corresponding
antifields, may have different weights. In this case the antifield
contribution to the abelian charges depends on the choice of operator 
$A$.
\item
One may consider the truncation to $p$-jets as a kind of regularization,
with the important property that it respects all Noether symmetries.
The true theory should then be recovered in the limit $p\to\infty$.
Unfortunately, the abelian charges diverge in this limit. This is
because the contributions from the fields and the antifields cancel to 
leading order in $p$, so the abelian charges are dominated by the
Noether antifields. In the absense of supersymmetry, these are all 
bosonic, and hence yield contributions of the same sign.
\end{enumerate}
These problems are annoying but I believe that they can be overcome.
At any rate, the finite $p$ theories are well-defined $DRO(N)$ modules,
parametrized by two data: the action $S$ and the integer $p$.

As argued above, my procedure produces quantum theories, at least in
the sense that Noether symmetries are represented projectively.
However, it is certainly not the orthodox approach to quantum mechanics.
Hence it is a pressing problem to make contact with standard quantum
theory. Of course, one can then not demand full $diff(N)$ symmetry, 
since quantum field theory is only well defined in flat space, with
a flat background metric transforming as a tensor field.
This connection is missing at the moment.

\section{What's wrong with string theory?}
There are several competitor theories of quantum gravity, most notably
superstring theory, and its recent incarnation M-theory, which has 
enjoyed the full attention of many prominent physicists since its 
resurrection around 1984 \cite{GSW86}. 
This might be somewhat surprising since the
only predictions that has come out of it (e.g., spacetime has 10 or 11 
dimensions, all particles have supersymmetric partners, and physics 
is governed by a gauge group with 496 generators) are in sharp 
disagreement with present-day experiments. 

In \cite{Lar97b}, I criticized string theory from an algebraic point of 
view, which is connected to the DRO algebra. Since the argument seems
to have passed unnoticed, and I still believe that it is relevant,
it is repeated here. The point is that a super-DRO algebra (including
the abelian extension) can be 
defined in a spacetime with has also fermionic directions. 
Hence the algebra $diff(N|M)$ of diffeomorphisms in a spacetime
with $N$ bosonic and $M$ fermionic directions admits abelian but
non-central extensions. 
Of course, this property is inherited
by every subalgebra of $diff(N|M)$, but in some (almost trivial) cases,
the generically non-central extension reduces to a central one. The
superconformal algebra is such an almost trivial case.

Thus, if we only consider central extension, the situation can be 
depicted by the diagram
\be
\begin{array}{ccccccc}
SC(0) & \into & diff(1|1) & \into & diff(N|M) & \into & \ldots \cr
\downarrow \hbox{c.e.} \cr
SC(c)
\end{array}
\label{scimb}
\ee
where $SC(c)$ denotes the superconformal algebra with central charge $c$
and the vertical arrow indicates central extension. From this point
of view, the superconformal algebra appears exceptionally deep; 
it is one of the few super Lie algebras admitting a central extension.
For a classification of centrally extended superalgebras see
\cite{GLS97}.
If we now allow for general abelian extensions, the diagram (\ref{scimb})
is replaced by
\be
\begin{array}{ccccccc}
SC(0) & \into & diff(1|1) & \into & diff(N|M) & \into & \ldots \cr
\downarrow \hbox{a.e.} && \downarrow \hbox{a.e.} && 
\downarrow \hbox{a.e.} \cr
SC(c) & \into & \tdiff(1|1) & \into & \tdiff(N|M) & \into & \ldots \cr
\end{array}
\ee
where the squares commute and $\tdiff(N|M)$ is a gauge-fixed version 
of the super-DRO algebra \cite{Lar97b}.
Now the superconformal algebra no longer appears exceptionally deep;
on the contrary, it is an unusually shallow structure in the sense that
the abelian extension is central. My criticism of string theory thus
boils down to this:

\medskip
{\em Why should a Theory of Everything be based on an unusually
shallow algebraic structure?}
\medskip

Lest it appears that I am critical only about string theory, I
must add that no other quantization procedure treats the quantum
diffeomorphism symmetry in a correct way, i.e. as a consistent Lie
algebra extension of the classical algebra.
In standard canonical quantization of gravity (ADM Hamiltonian), 
the constraints do not even classically reproduce the diffeomorphism 
algebra, but only the so-called ``Dirac algebra''. As noted e.g. in 
\cite{Ish91}, page 169, this is not quite $diff(N)$; in fact, it is
not even a proper Lie algebra.

\section{Three-dimensional critical phenomena}
The main thesis of this paper is that the $DRO(N)$ representation 
theory is useful for understanding quantum gravity. However, I believe
that it can also be important for another application in physics, namely
the classification of three-dimensional critical phenomena. Indeed,
it was this problem that originally arose my interest in the
diffeomorphism algebra. 

The key observation is that critical exponents are {\em universal};
two similar statistical models do not just have similar, but exactly
the same (or completely different) critical exponents. Although
universality is not strictly proven, it can be justified with
renormalization group arguments. Conformal field theory (CFT)
\cite{BPZ84, FMS96} gives another explanation of two-dimensional 
universality: all fields must form representations
of the conformal algebra $Vir\oplus{Vir}$ (or some more complicated
chiral algebra), and these typically fall into a discrete series.
Since universality is not restricted to two dimensions, one may hope 
that there is some similar explanation in higher dimensions.
The conformal algebra will not do; it does make some predictions, e.g.
about the form of the three-point function, but it does not predict the
spectrum of critical exponents. The idea is to instead look for a 
larger algebra already in two dimensions, and generalize this to higher
dimensions. The diffeomorphism algebra is the only reasonable candidate.

Let $\hh=Vir\oplus{Vir}$ be the conformal algebra in two dimensions,
and let $\oj = DRO(2)$. Then $\hh\subset\oj$, with the standard 
embedding: $z=x^0+ix^1$, $\bar z = x^0-ix^1$, and
the central charges are given by $c = \bar c = 12c_1$ if $c_2=0$. 
(If $c_2$ is non-zero, the situation is more complicated because the
cross bracket $[L_m,\bar L_n]$ acquires a non-central extension.)
Given any unitary $\oj$ module, we obtain a unitary $\hh$ module by
restriction. Conversely, given a unitary $\hh$ module $\rep$, we can
always construct a $\oj$ module, e.g. the induced module
${\rm Ind}^\oj_\hh(\rep)$ (recall that the induced module is the
universal enveloppe of $\oj$, modulo the ideal generated by the
representation $\rep$. However, unitarity is not guaranteed in
this case. 

We have thus shown that $DRO(2)$ contains at least as much information
as the conformal algebra, and thus predicts the same spectrum of
critical exponents as the latter. This immediately suggest the
generalization to higher dimensions: critical exponents in $N$ 
dimensions should follow from the dilatation eigenvalues in 
unitary irreps of $DRO(N)$. Unfortunately, the classification of such
modules, and even the construction of the simplest examples, is
presently beyond my ability. Note that this argument depends solely
on the existence of abelian extensions, whose restriction to the 
conformal algebra is the central Virasoro extension. Hence, it could
have been made already when the first $diff(N)$ extensions were found
in 1989 \cite{Lar89}.

\end{document}